\documentclass[referee]{raa}            

\usepackage{graphicx,times}    
\usepackage{natbib}
\usepackage{amssymb,amsmath}
\bibpunct{(}{)}{;}{a}{}{,}
\usepackage{threeparttable}
\usepackage{float}
\usepackage{hyperref}

\begin{document}


\title{Mergers Drive Structural Complexity but Not Starbursts in 
Lyman-$\alpha$ Emitters at $3 < z < 4$: A JWST Spatially Resolved View
}


\volnopage{Vol.0 (20xx) No.0, 000--000}      
\setcounter{page}{1}          

\author{Qi Song \inst{1,2}
\and F. S. Liu \inst{1,2,3} \thanks{E-mail: fsliu@nao.cas.cn}
\and Jian Ren \inst{1,4} \thanks{E-mail: renjian@bao.ac.cn}
\and Tianfu Gao \inst{5} \thanks{E-mail: tianfugao@163.com}
\and Pinsong Zhao \inst{6,1}
\and Qifan Cui \inst{7,1}
\and Yubin Li \inst{8,1}
\and Hao Mo \inst{1,2,3}
\and Guanghuan Wang \inst{9,1}
    }

\institute{National Astronomical Observatories, Chinese Academy of Sciences, 20A Datun Road, Chaoyang District, Beijing 100101, China\\
\and
    Key Laboratory of Optical Astronomy, National Astronomical Observatories, Chinese Academy of Sciences, 20A Datun Road, Chaoyang District, Beijing 100101, China\\
\and
    School of Astronomy and Space Science, University of Chinese Academy of Science, Beijing 100049, China\\
\and
    Key Laboratory of Space Astronomy and Technology, National Astronomical Observatories, Chinese Academy of Sciences, 20A Datun Road, Chaoyang District, Beijing 100101, China\\
\and
    College of Physical Science and Technology, Shenyang Normal University, Shenyang 110034, China\\
\and 
    Kavli Institute for Astronomy and Astrophysics, Peking University, Beijing 100871, China\\
\and
    Shanghai Key Lab for Astrophysics, Shanghai Normal University, Shanghai 200234, China\\
\and
    College of Engineering and Technology, Baoshan University, Baoshan, Yunnan 678000, China\\
\and 
    Purple Mountain Observatory, Chinese Academy of Sciences, 10 Yuanhua Road, Nanjing 210034, China\\
\vs\no
   {\small Received 20xx month day; accepted 20xx month day}}

\abstract{
Recent observations with the James Webb Space Telescope (JWST) reveal that the merger fraction among Ly$\alpha$ emitters (LAEs) at redshifts $z > 3$ is significantly higher than previously estimated. In this study, we focus on three high signal-to-noise merging LAE systems at $3 < z < 4$, selected from the VLT/MUSE-Deep survey in the GOODS-S field. We combine new \textit{JWST}/NIRCam broadband and medium-band imaging with archival \textit{HST}/ACS data to perform spatially resolved spectral energy distribution (SED) fitting using the \textsc{Bagpipes} software package. Our analysis reveals that two of the systems are minor mergers, while the third is a major merger. The close agreement between spatially resolved and integrated stellar mass estimates indicates that recent star formation does not significantly outshine the light from older stellar populations in these systems. Moreover, both the individual components and the systems as a whole lie on the star-forming main sequence, further supporting the conclusion that these mergers have not yet triggered substantial starburst activity. 
Furthermore, we detect prominent color gradients and disturbed dust distributions in these merging systems, indicating that the mergers have already induced significant internal structural perturbations. These morphological and dust-related changes may facilitate the escape of Ly$\alpha$ photons -- potentially through mechanisms such as gas redistribution or a reduced covering fraction of neutral hydrogen -- thereby playing a key role in shaping the observed properties of LAEs.
}
   \authorrunning{Q. SONG ET AL.}            
   \titlerunning{SPATIALLY RESOLVED ANALYSIS OF LAE MERGERS}  

   \maketitle

%
%
\section{Introduction}\label{sect:intro}

Since the pioneering prediction by \citet{Partridge67} that Lyman-alpha (Ly$\alpha$) emission could serve as a powerful tracer of galaxy formation and evolution in the early universe -- and the subsequent discovery of the first Ly$\alpha$ emitters (LAEs) \citep{Hu1996Natur.382..231H,Pascarelle1996Natur.383...45P,Cowie1998AJ....115.1319C} -- Ly$\alpha$ radiation has become a cornerstone probe for studying the physical properties of high-redshift galaxies and the epoch of cosmic reionization. This is due to the resonant scattering of Ly$\alpha$ photons by neutral hydrogen, which renders them exquisitely sensitive to the conditions in both the interstellar medium (ISM) and the intergalactic medium (IGM) \citep{Hayes2011b,Dijkstra14,zhang24}.

Advances in ground-based telescopes (e.g., Keck, VLT, Subaru) and the \textit{Hubble Space Telescope} (\textit{HST}) have enabled the identification of large samples of LAEs \citep{Taniguchi05,Gronwall07,Ouchi08,Ouchi10,Nakajima12,Matthee15,Herenz2017,Pentericci18,Ning2020ApJ...903....4N,Ning2022ApJ...926..230N,Bacon23,Talia2023A&A...678A..25T}. These datasets provide a robust foundation for tracing the evolution of LAEs across cosmic time. In particular, they have facilitated systematic studies of their luminosity functions \citep{Ouchi08,Ouchi10,Cowie2010ApJ...711..928C,Konno2016ApJ...823...20K,Konno2018PASJ...70S..16K,Ning2022ApJ...926..230N,Umeda2025ApJS..277...37U}, morphological properties \citep{Ouchi10,Bond2012ApJ...753...95B,Kobayashi2016ApJ...819...25K,Paulino-Afonso2018MNRAS.476.5479P,Shibuya2019ApJ...871..164S}, and interstellar medium (ISM) characteristics \citep{Finkelstein2011ApJ...733..117F,nakajima2014ionization,Konno2016ApJ...823...20K}, offering critical insights into the formation mechanisms, physical conditions, and internal structures of galaxies in the early universe.

With the successful launch and operation of the James Webb Space Telescope (JWST), studies of high-redshift LAEs have entered a transformative new era. JWST's unprecedented sensitivity in the near- and mid-infrared, combined with its high spatial resolution \citep{Rieke2023ApJS..269...16R, Rigby2023PASP..135d8001R}, enables the direct detection of rest-frame optical and near-infrared emission from galaxies at $z > 3$. This capability permits significantly more accurate measurements of their physical properties -- such as stellar mass, star formation rate, and metallicity -- than were previously possible \citep{Yang2022ApJ...938L..17Y, Morishita2024ApJ...963....9M, Miller2024arXiv241206957M, Allen2025A&A...698A..30A, Yang2025arXiv250407185Y}. It overcomes key limitations of earlier instruments, which were largely restricted to rest-frame ultraviolet observations and hampered by lower spatial resolution.

Previous studies have generally suggested that high-redshift LAEs are young ($\sim 1$--100~Myr), low-mass ($M_* \sim 10^8$--$10^9\,M_\odot$), and compact galaxies ($r_e \lesssim 1$~kpc) with low dust content ($E(B-V) \sim 0$--0.2) \citep{Ono10,Nakajima12,Kojima17,Paulino-Afonso2018MNRAS.476.5479P}, and that these properties are closely linked to the production and escape of Ly$\alpha$ photons \citep{Malhotra12}. However, recent \textit{JWST} observations have revealed a significant population of merging systems among LAEs \citep{Witten2024NatAs...8..546W,Liu2024ApJ...966..210L,Ning2024ApJ...963L..38N}. More recently, \citet{Ren25}, based on \textit{JWST} imaging of LAEs in the GOODS-S field, reported a merger fraction substantially higher than earlier estimates. 
If mergers are indeed more common among LAEs, our understanding of Ly$\alpha$ photon production and escape mechanisms in such systems may need revision--particularly because prior interpretations largely assumed LAEs to be predominantly isolated, compact galaxies. Galaxy mergers can induce strong dynamical disturbances, gas inflows/outflows, and localized enhancements in star formation, all of which alter the distribution and kinematics of ionized gas and thereby modify the scattering and escape pathways of Ly$\alpha$ photons 
\citep{2011Natur.476..304H,Cabot16,Chang23}. Moreover, merger-driven turbulence and stellar feedback within the interstellar medium (ISM) may carve out low-density channels that further facilitate Ly$\alpha$ escape \citep{Witten2024NatAs...8..546W}.
Consequently, spatially resolved studies of merging LAEs are crucial for reexamining the connections between Ly$\alpha$ emission, galaxy morphology, and kinematics.

In this work, we investigate three high signal-to-noise merging LAE systems at $3 < z < 4$, selected from the VLT/MUSE-Deep survey in the GOODS-S field. We combine new \textit{JWST}/NIRCam broadband and medium-band imaging with archival \textit{HST}/ACS data to perform a spatially resolved analysis. Using the \textsc{Bagpipes} software package \citep{Bagpipes18}, we carry out pixel-by-pixel spectral energy distribution (SED) fitting to construct two-dimensional maps of key physical properties for each component, including stellar mass, star formation rate (SFR), mass-weighted age, V-band attenuation ($A_V$), specific star formation rate (sSFR), and UV continuum slope ($\beta$). 
Our results show that all individual components and the systems as a whole lie on the star-forming main sequence, with stellar masses from spatially resolved and integrated measurements in excellent agreement--indicating that these mergers have not yet triggered significant starburst activity. Strong color variations and disturbed dust distributions reveal spatial inhomogeneities in both star formation and extinction, reflecting substantial internal structural perturbations already induced by the mergers. Such morphological and dust--related changes may facilitate Ly$\alpha$ photon escape -- potentially through mechanisms such as gas redistribution or a reduced covering fraction of neutral hydrogen -- thereby playing a key role in shaping the observed properties of LAEs.

The structure of this paper is as follows. Section~\ref{2} describes the \textit{JWST} data used in this study and the selection of the 
LAE merger sample; Section~\ref{3} outlines the methodology for defining effective pixels and performing spectral energy distribution (SED) fitting; Section~\ref{4} presents our results; Section~\ref{5} discusses their implications; and Section~\ref{6} summarizes our main conclusions. Throughout this work, we adopt a concordance cosmology with $H_0 = 70~\mathrm{km\,s^{-1}\,Mpc^{-1}}$, $\Omega_{\rm m} = 0.3$, and $\Omega_\Lambda = 0.7$.

\section{Data and Sample}\label{2}
\label{sect:2}

\subsection{JWST data}\label{sect:2.1}

This work is based on imaging data from the \textit{JWST}-SPRING program (\textit{Spatially Pixel-level Resolved Investigations into Nascent Galaxies with the James Webb Space Telescope}).  
The program delivers homogenized, pixel-level \textit{JWST} imaging covering all major deep extragalactic survey fields previously observed by the \textit{Hubble Space Telescope} (\textit{HST}), including the five CANDELS regions (COSMOS, EGS, GOODS-S, GOODS-N, and UDS), the extended COSMOS field, and numerous smaller ancillary fields—collectively spanning nearly 1 square degree of sky.  
\textit{JWST}-SPRING provides extensive, uniform, and deep multi-wavelength imaging from \textit{JWST}'s NIRCam and MIRI instruments, complemented by spectroscopic data from NIRSpec and NIRCam/WFSS.  
These datasets are carefully aligned with archival \textit{HST} imaging (UVIS, ACS, and WFC3/IR) and slitless spectroscopy, enabling a wide range of pixel-scale extragalactic studies. 

In this work, we primarily use the \textit{JWST}/NIRCam imaging data in the GOODS-S field. 
The multi-band raw \textit{JWST}/NIRCam imaging data for GOODS-S -- including the broadband filters F090W, F115W, F150W, F200W, F277W, F356W, and F444W, as well as the medium-band filters F182M, F210M, F335M, and F410M -- were drawn from Programs 1176, 1180, 1210, 1283, 1286, 1287, 1895, 2079, 2198, 2514, 2516, 3215, 3954, 3990, 6434, 6511, and 6541. 

To preserve finer image details and enable joint analysis with HST data, we drizzled all JWST images to a uniform pixel scale of $0.03''\,\mathrm{pixel}^{-1}$, matching the native resolution of the shorter-wavelength JWST/NIRCam bands. This processing step enhances spatial consistency and retains higher-resolution structural information, thereby facilitating more reliable pixel-by-pixel SED fitting for the galaxies.
A detailed description of the data processing pipeline, calibration procedures, and full documentation is available on the \textit{JWST}-SPRING website\footnote{\url{http://groups.bao.ac.cn/jwst_spring/}} and in the program overview by Liu et al.\ (in preparation).

\subsection{Sample Selection} \label{sect:2.2}

The LAE sample is drawn from the catalog presented in our previous work \citep{Ren25,Song25}.  
Given the requirement for multi-band imaging in our subsequent analysis, we restrict the sample to sources identified in the MUSE-Deep spectroscopic survey \citep{Bacon17,Inami17,Bacon23}. MUSE-Deep is a deep integral-field spectroscopic program carried out with the Multi Unit Spectroscopic Explorer (MUSE; \citealt{Bacon2010}) on the Very Large Telescope (VLT). Its primary objective is to probe the physical mechanisms driving galaxy formation and evolution in the high-redshift universe using three-dimensional spectroscopic data.  
We utilize data from the second data release of MUSE-Deep (\citealt{Bacon23}, hereafter DR2), which includes a comprehensive reprocessing of the earlier first data release (\citealt{Bacon17,Inami17}, hereafter DR1). DR2 comprises three distinct fields: a 10-hour $3 \times 3\,\mathrm{arcmin}^2$ mosaic (MOSAIC), a 31-hour $1 \times 1\,\mathrm{arcmin}^2$ region (UDF-10), and the 141-hour MUSE eXtremely Deep Field (MXDF), which covers a circular area of 1 arcmin in diameter.  In our prior work, candidate LAEs were selected using the following criteria: redshift confidence flag \texttt{ZCONF = 2} or \texttt{3} (corresponding to Ly$\alpha$ line signal-to-noise ratios of at least 5 and 7, respectively) and an additional requirement of Ly$\alpha$ S/N $>$ 3. 

We restrict the LAE sample to the redshift range $3 < z < 4$, ensuring rest-frame spectral coverage from the far-ultraviolet to the near-infrared -- 
essential for robust estimation of physical parameters. Cross-matching with the merger catalog of \citet{Ren25} yields 127 LAEs classified as merger candidates within this redshift interval. To focus on systems amenable to detailed morphological analysis, we further restrict our sample to merger candidates with relatively simple structures. 
We first identify dual-nucleus systems by visually inspecting the JWST/NIRCam F200W images, which offer high spatial resolution and sample the rest-frame optical continuum. A pair of nuclei is deemed clearly distinguishable if each exhibits a local surface-brightness peak in at least two adjacent bands (e.g., F150W and F200W) and the profile between them displays a distinct local minimum -- indicative of a double-nucleus morphology. We then measure the angular separation between the two peaks in the F200W image using \texttt{Photutils} \citep{photutils24}. Given that Ly$\alpha$ emitters are typically compact \citep{Shibuya2019ApJ...871..164S,Song25}, merging nuclei are expected to be closely spaced. We therefore select only systems with projected separations between $0.15''$ and $0.30''$ (corresponding to 5--10 pixels at our pixel scale of $0.03''\,\mathrm{pixel}^{-1}$). This range ensures the nuclei are close enough to be physically interacting -- likely undergoing a merger -- yet sufficiently separated to be resolved by JWST’s diffraction-limited resolution. Applying these criteria yields a final sample of 26 candidate 
dual-nucleus systems.
For this subsample, we require secure detections in all \textit{JWST}/NIRCam broad-band and medium-band filters as well as in the \textit{HST}/ACS bands (F435W, F606W, F775W, and F814W), with the latter providing critical constraints on the rest-frame UV continuum. To ensure high-fidelity spectral energy distribution (SED) fitting, we impose a signal-to-noise ratio (S/N) threshold of $>15$ in all these bands. Applying these selection criteria, we identify three LAE merger candidates with unambiguous multicomponent morphologies: MD-6666, MD-82, and MD-1711.

\begin{figure*}[!t]
\centering
\includegraphics[width=1\textwidth]{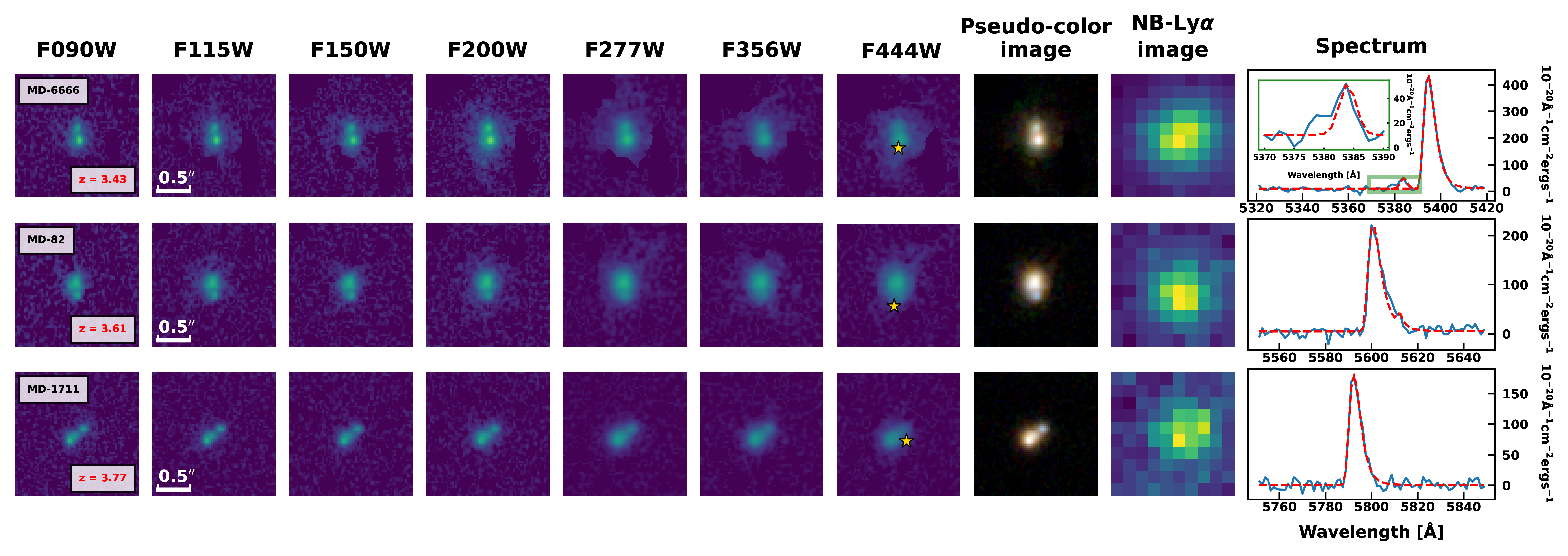} 
\caption{
From left to right: JWST/NIRCam broadband images in F090W, F115W, F150W, F200W, F277W, F356W, and F444W; 
pseudo-color composite (blue: F090W, F115W, F150W; green: F200W, F277W; red: F356W, F444W); 
MUSE Ly$\alpha$ narrow-band image and the emission-line spectrum with an asymmetric Gaussian fit. All panels show a $2'' \times 2''$ field of view. Yellow stars mark the Ly$\alpha$ surface-brightness peaks in the narrow-band map and are overlaid on the F444W image.}
\label{fig:Sample_image}
\end{figure*}

Table~\ref{tab:table1} summarizes the key properties of the three LAE merger systems. Rest-frame Ly$\alpha$ equivalent widths and luminosities are taken from the catalog of \citet{Bacon23}. Figure~\ref{fig:Sample_image} presents \textit{JWST}/NIRCam broadband imaging, MUSE-derived narrowband maps, and spectroscopic data for these three spectroscopically confirmed LAE mergers. Imaging in all \textit{HST}/ACS and \textit{JWST}/NIRCam bands is shown in Appendix~\ref{fig:mul_image}.

We model the Ly$\alpha$ emission line profiles extracted from the MUSE spectra. To account for complex gas kinematics and attenuation by the intergalactic medium (IGM), we adopt the asymmetric Gaussian profile introduced by \citet{Shibuya14} to describe the Ly$\alpha$ line shape. This profile is defined as

\begin{equation} \label{eq:Lya_fitting}
    f(\lambda) = A \exp\left(-\frac{\left(\lambda - \lambda_{0}^{\text{asym}}\right)^{2}}{2\sigma_{\text{asym}}^{2}}\right) + f_{0},
\end{equation}

where $A$ is the line amplitude, $\lambda_0^{\text{asym}}$ is the peak wavelength, and $f_0$ represents the underlying continuum level. The dispersion $\sigma_{\text{asym}}$ is wavelength-dependent and given by 
\begin{equation}
\sigma_{\mathrm{asym}} = a_{\mathrm{asym}}(\lambda - \lambda_0^{\mathrm{asym}}) + d.
\label{eq:sigma_asym}
\end{equation}
Here, $a_{\mathrm{asym}}$ is the linear coefficient that governs the wavelength dependence of the Gaussian standard deviation. It is defined mathematically as the first derivative

\begin{equation}
a_{\mathrm{asym}} = \frac{\partial \sigma_{\mathrm{asym}}}{\partial (\lambda - \lambda_0^{\mathrm{asym}})}.
\label{eq:asym_coefficient}
\end{equation}

The parameter $d$ represents the baseline width at $\lambda_0^{\mathrm{asym}}$. 
The sign of $a_{\mathrm{asym}}$ determines the direction of line asymmetry: $a_{\mathrm{asym}} > 0$ corresponds to broadening of the redshifted (red) wing, whereas $a_{\mathrm{asym}} < 0$ indicates broadening of the blueshifted (blue) wing.

\begin{table}[ht]
\centering
\begin{threeparttable}
\caption{Basic properties of three LAE merger candidates}
\label{tab:table1}
\small
\begin{tabular}{lccccccc}
\hline
ID & R.A. & Decl. & $z_{\rm spec}$ & zconf & EW$_{0}$ & log($L_{\rm Ly\alpha}$) & LINE\_SNR\_MAX\tnote{a} \\
   & [deg] & [deg] &  &  & [\AA] & [erg/s] &  \\
\hline
MD-6666   & 53.1595  & -27.7767 & 3.43 & 3 & 49.85 & 42.46 & LYALPHAb\tnote{b} \\
MD-82     & 53.1516  & -27.7853 & 3.61 & 2 & 30.28 & 42.18 & LYALPHA \\
MD-1711   & 53.1416  & -27.7749 & 3.77 & 3 & 59.49 & 42.25 & LYALPHA \\
\hline
\end{tabular}
\begin{tablenotes}
\footnotesize
\item[a] Name of the emission or absorption line with the highest signal-to-noise ratio.
\item[b] The suffix ``$b$'' appended to a line name denotes a blended feature \citep{Bacon23}.
\end{tablenotes}
\end{threeparttable}
\end{table}

\section{Methods}\label{3}

\subsection{Determination of effective pixels}\label{3.1}

Our analysis utilizes imaging from \textit{JWST}/NIRCam (F090W, F115W, F150W, F182M, F200W, F210M, F277W, F335M, F356W, F410M, F444W) and \textit{HST}/ACS (F435W, F606W, F775W, F814W) for SED fitting, enabling a comprehensive characterization of the stellar populations in these systems.  
Although \textit{HST}/WFC3 (F105W, F125W, F140W, F160W) provides overlapping wavelength coverage, we exclude these data from our analysis due to the superior spatial resolution of \textit{JWST}/NIRCam. This broad wavelength baseline -- spanning the rest-frame far-ultraviolet to near-infrared at $3 < z < 4$ -- allows us to tightly constrain key physical parameters, including stellar mass, star formation rate (SFR), dust attenuation ($A_V$), and mass-weighted age.

Prior to SED fitting, we homogenized the point-spread functions (PSFs) of all \textit{JWST}/NIRCam and \textit{HST} images using products from the \textit{JWST}-SPRING pipeline. The PSFs were constructed from 114 isolated stars in the GOODS-S field using the \texttt{PSFr} Python package \citep{psfr}.  
Details on the stellar selection criteria and software configuration are provided in Cui et al.\ (in preparation). 
We derived convolution kernels with \texttt{PyPHER} \citep{pypher} to match all PSFs to that of the F444W band, which has the broadest PSF among the selected filters (FWHM = 0.163$''$) and thus serves as the common resolution reference.  
All images were convolved accordingly to ensure consistent spatial resolution, minimizing PSF-induced biases in multi-band light profiles during SED fitting. 
To define effective pixel regions for each source, we adopted the methodology of \citet{Clara23}, applying a signal-to-noise ratio (S/N) threshold of 5 to the F150W and F200W bands. 
This yields between 215 and 287 effective pixels per source, each corresponding to a physical scale of $\sim$220 pc at $z \sim 3.5$.  
Pixels falling below this S/N threshold were excluded from all imaging products and quantitative measurements. 
Finally, we cropped all science images to these effective pixel masks and used \texttt{Photutils} to decompose the merger components, isolating individual galaxy segments within each interacting system for precise, component-resolved analysis. 

Specifically, we applied the \texttt{deblend\_sources} function to the JWST/NIRCam F200W images using the parameters \texttt{npixels=5}, \texttt{nlevels=32}, and \texttt{contrast=0.1}. Here, \texttt{nlevels} specifies the number of multi-thresholding levels, exponentially spaced between a source’s minimum and maximum pixel values, enabling fine separation of closely spaced bright cores. This automated decomposition was applied only to systems that exhibit clear double-nucleus structures in at least two adjacent bands, and all results were visually inspected to ensure physical plausibility.

\subsection{SED fitting}\label{3.2}

This work employs the \textsc{Bagpipes} (Bayesian Analysis of Galaxies for Physical Inference and Parameter EStimation; \citealt{Bagpipes18}) software package for all spectral energy distribution (SED) fitting, encompassing both integrated-galaxy and pixel-by-pixel analyses.  
\textsc{Bagpipes} is a Bayesian spectral modeling code designed to reproduce galaxy emission across a wide wavelength range -- from the far-ultraviolet to the microwave -- by combining stellar population synthesis with prescriptions for dust attenuation, nebular emission, and star formation histories. 
The code fits these models to arbitrary combinations of photometric and spectroscopic data using the nested sampling algorithm of \citet{Skilling2006}, as implemented in \textsc{MultiNest} \citep{Feroz08,Feroz09,Feroz19} via the \texttt{PyMultiNest} Python interface \citep{Buchner2014}. This approach efficiently explores high-dimensional parameter spaces, yielding both posterior probability distributions and Bayesian evidence for model comparison. 
Thanks to its flexible and physically motivated framework, \textsc{Bagpipes} is especially well suited for complex SED modeling tasks, including spatially resolved analyses of interacting systems. To minimize systematic biases, we apply identical parameter configurations uniformly across all fitting instances--whether for entire galaxies or individual pixels. This consistent setup ensures that any variations in the inferred physical properties (e.g., stellar mass, star formation rate, dust attenuation) among merger components reflect genuine astrophysical differences rather than artifacts of inconsistent modeling choices.

To characterize differential physical properties across merger components, we adopt a double-power-law star formation history (SFH) model. This parametrization assumes that the star formation rate (SFR) rises and declines independently, enabling asymmetric SFHs commonly observed in dynamically evolving systems such as galaxy mergers \citep{Bagpipes18,bagpipes19}. Although computationally demanding, this flexible form significantly improves the model’s ability to capture complex star formation episodes compared to traditional exponentially declining or delayed-$\tau$ models. In particular, the double-power-law SFH more accurately represents both early quiescent formation and recent starbursts--features expected in merger-driven star formation scenarios \citep{Bagpipes18}. The model is mathematically defined as:
\begin{equation} \label{eq:double_power_law}
    \mathrm{SFR}(t) \propto 
    \left[
        \left( \frac{t}{\tau} \right)^{\alpha} + 
        \left( \frac{t}{\tau} \right)^{-\beta}
    \right]^{-1},
\end{equation}
where $\alpha$ and $\beta$ control the falling and rising slopes of the SFH, respectively, and $\tau$ sets the timescale at which star formation peaks \citep{Bagpipes18}. We assign logarithmic priors to both slopes, allowing $\alpha, \beta \in [-3,\, 3]$. The peak time parameter $\tau$ is permitted to vary between 1~Myr and 15~Gyr, while metallicity is allowed to range from 0 to $Z_{\odot}$. The total formed stellar mass is sampled over the logarithmic interval $\log(M/M_{\odot}) \in [5,\, 13]$. 
We use synthetic stellar population templates from \citet{BC03}, assuming a \citet{Kroupa} initial mass function (IMF). Nebular emission (including both continuum and line emission) is modeled self-consistently using \textsc{Cloudy} \citep{Ferland17}, with the ionization parameter $\log U$ uniformly sampled in the range $[-4,\, -1]$. Dust attenuation follows the \citet{Calzetti00} law, with the visual extinction $A_V$ allowed to vary within $0 < A_V < 2$.

\section{RESULTS}\label{4}

\subsection{Are These Systems Genuine Mergers ?}\label{4.1}

While two of our LAE merger systems -- MD-6666 and MD-82 \citep{Vitte25} -- exhibit double-peaked Ly$\alpha$ line profiles, 
such features alone are insufficient to confirm physical mergers due to the complex radiative transfer of Ly$\alpha$ photons in neutral gas. 
To robustly establish their merger nature, we performed SED fitting on each galaxy component identified in Section~\ref{3.1}. Integrated photometry for each component was obtained by summing the flux over all constituent pixels.  
The SED fitting was carried out using the \textsc{Bagpipes} framework described in Section~\ref{3.2}, with redshift treated as a free parameter.  
Allowing redshift to vary enables a direct comparison between photometric and spectroscopic redshifts, providing an independent validation of the component associations -- particularly important in systems with complex or ambiguous morphologies.

Table~\ref{table2} summarizes the final SED fitting results. 
Stellar mass, mass-weighted age, dust attenuation ($A_V$), star formation rate (SFR), and photometric redshift ($z_{\mathrm{phot}}$) 
are reported as the 50th percentiles of the \textsc{Bagpipes} posterior distributions, 
with uncertainties corresponding to the 16th and 84th percentiles (i.e., $\pm1\sigma$ confidence intervals). 
The rest-frame UV continuum slope, $\beta$, was derived by fitting a power law to the median model spectrum from the \textsc{Bagpipes} 
posterior over the rest-frame wavelength range 1268--2580~\AA\ \citep{Calzetti1994}:
\begin{equation} \label{eq:beta}
    F_{\lambda} \propto \lambda^{\beta}.
\end{equation}
The specific star formation rate (sSFR) was computed directly from the median values of the SFR 
and stellar mass obtained from the \textsc{Bagpipes} posteriors.

\begin{table}[ht]
\centering
\caption{Physical properties of each component in three LAE merger systems}
\label{table2}
\footnotesize
\begin{tabular}{lccccccccc}
\hline
ID & $\log{M_\ast/M_\odot}$ & $A_{V}$ & Age & SFR & $z_{\rm phot}$  & $z_{\rm spec}$ & UV slope $\beta$ & sSFR \\
   &                 & [mag] & [Myr] & [$M_\odot$/yr] &  &  &  & [Gyr$^{-1}$] \\
\hline
MD-6666-N  & 7.64$_{-0.11}^{+0.12}$ & 0.65$_{-0.05}^{+0.05}$ & 9.52$_{-3.0}^{+5.5}$  & 0.49$_{-0.12}^{+0.17}$ & 3.42$_{-0.02}^{+0.01}$  & 3.43 & -1.67$_{-0.03}^{+0.03}$ & 11.1$_{-0.4}^{+0.3}$ \\
MD-6666-S  & 7.98$_{-0.11}^{+0.12}$ & 0.47$_{-0.04}^{+0.04}$ & 11.16$_{-3.7}^{+6.6}$  & 1.06$_{-0.26}^{+0.40}$ & 3.42$_{-0.01}^{+0.01}$  & 3.43 & -1.89$_{-0.03}^{+0.03}$ & 11.2$_{-0.3}^{+0.3}$ \\
MD-82-N    & 9.00$_{-0.07}^{+0.06}$ & 0.11$_{-0.06}^{+0.07}$ & 393.18$_{-113.2}^{+124.0}$ & 1.96$_{-0.23}^{+0.30}$ & 3.70$_{-0.03}^{+0.03}$  & 3.61 & -2.08$_{-0.03}^{+0.03}$ & 2.0$_{-0.1}^{+0.1}$ \\
MD-82-S    & 7.97$_{-0.15}^{+0.09}$ & 0.14$_{-0.08}^{+0.08}$ & 128.34$_{-38.0}^{+51.1}$ & 0.52$_{-0.08}^{+0.09}$ & 3.70$_{-0.04}^{+0.03}$  & 3.61 & -2.35$_{-0.02}^{+0.02}$ & 5.5$_{-0.2}^{+0.3}$ \\
MD-1711-NE  & 7.19$_{-0.12}^{+0.14}$ & 0.33$_{-0.07}^{+0.08}$ & 11.17$_{-4.8}^{+12.1}$  & 0.17$_{-0.04}^{+0.08}$ & 3.77$_{-0.02}^{+0.02}$  & 3.77 & -2.03$_{-0.03}^{+0.03}$ & 11.2$_{-0.4}^{+0.6}$ \\
MD-1711-SW  & 8.19$_{-0.16}^{+0.11}$ & 0.37$_{-0.12}^{+0.11}$ & 48.64$_{-24.8}^{+41.4}$  & 1.55$_{-0.30}^{+0.25}$ & 3.77$_{-0.03}^{+0.03}$  & 3.77 & -1.92$_{-0.03}^{+0.03}$ & 10.0$_{-1.1}^{+1.7}$ \\
\hline
\end{tabular}
\end{table}

\begin{figure*}[t]
\centering
\includegraphics[width=1\textwidth]{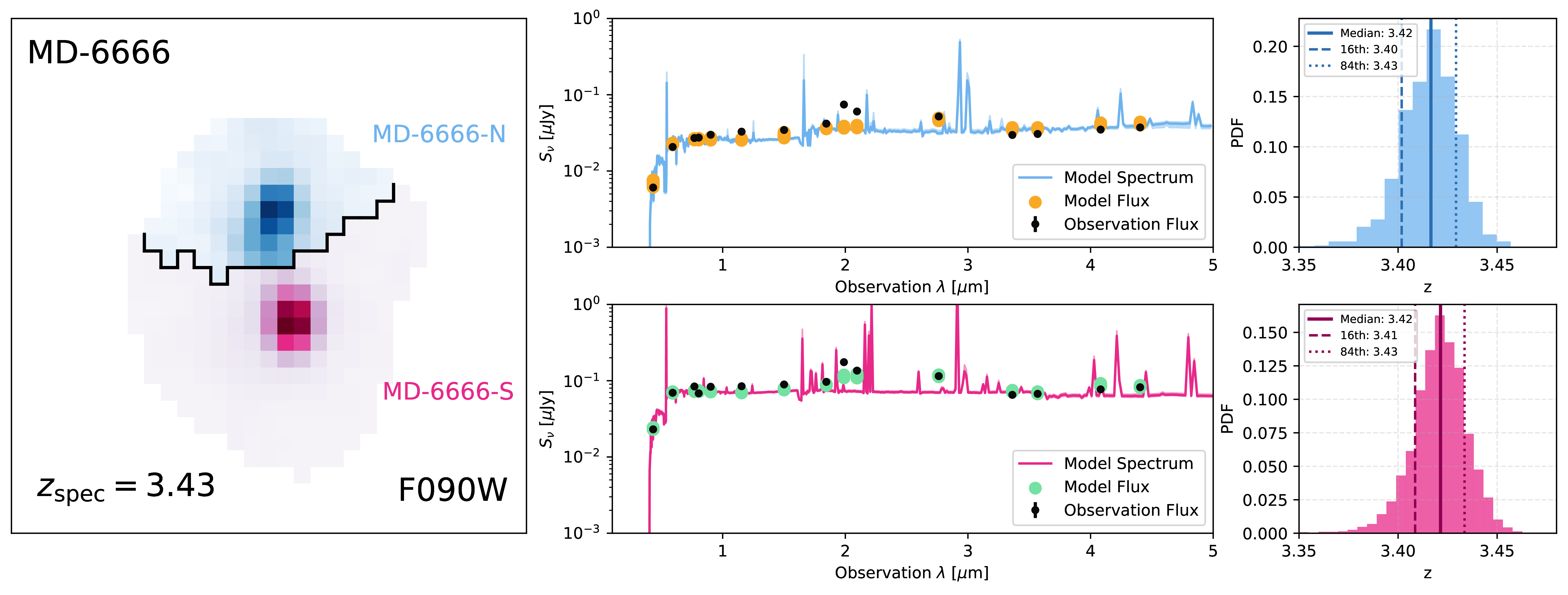} 
\caption{
Illustration of the SED fitting results for the galaxy components in the LAE merger system MD-6666 at $z = 3.43$. 
The left panel shows the JWST/NIRCam F090W image, with MD-6666-N (upper component) and MD-6666-S (lower component) 
highlighted using distinct color maps. 
Black contours delineate the segmented regions used for photometry. 
The middle two panels display the \textsc{Bagpipes} SED fits for MD-6666-N and MD-6666-S, respectively, 
plotted on identical vertical scales to enable direct comparison. 
The right panels show the corresponding posterior redshift distributions from the SED fitting, with the median values and 16th -- 84th percentile credible intervals indicated.
}
\label{fig:6666}
\end{figure*}

\begin{figure*}[t]
\centering
\includegraphics[width=1\textwidth]{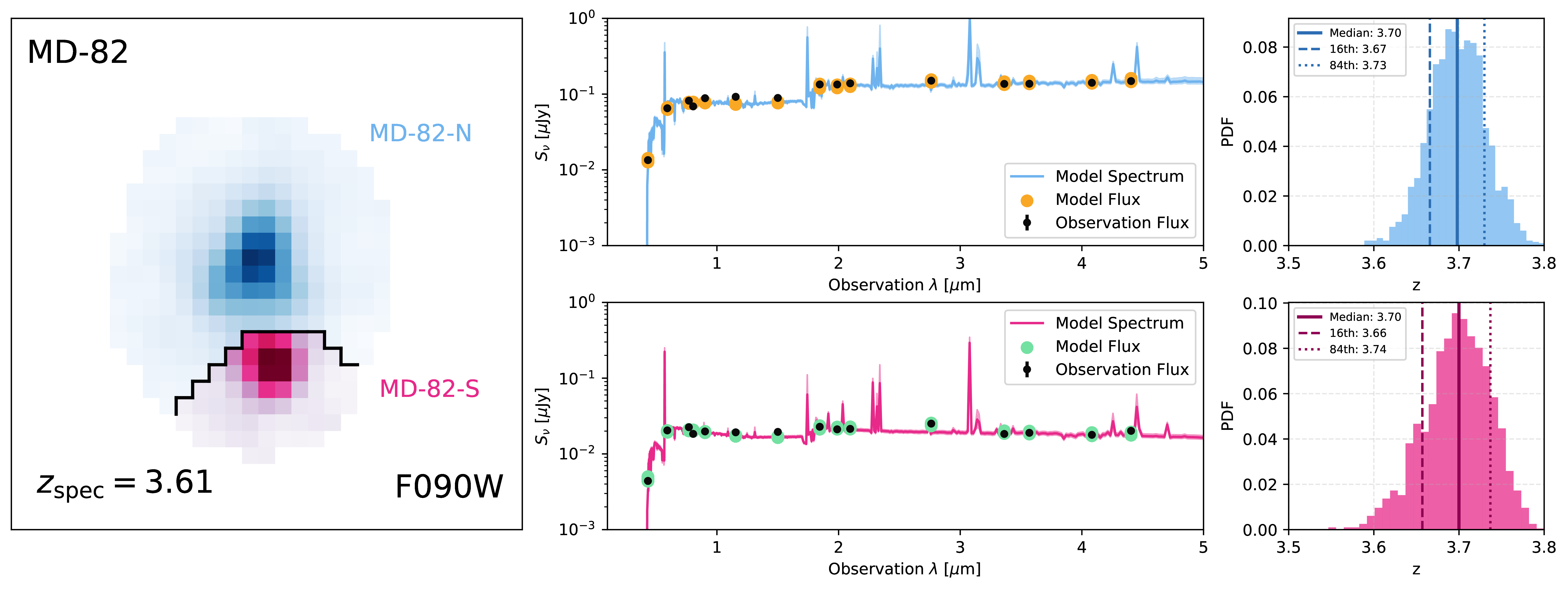} 
\caption{
Illustration of the SED fitting results for the LAE merger system MD-82 at $z = 3.61$. 
The symbols and layout follow the same convention as in Figure~\ref{fig:6666}.}
\label{fig:82}
\end{figure*}

\begin{figure*}[t]
\centering
\includegraphics[width=1\textwidth]{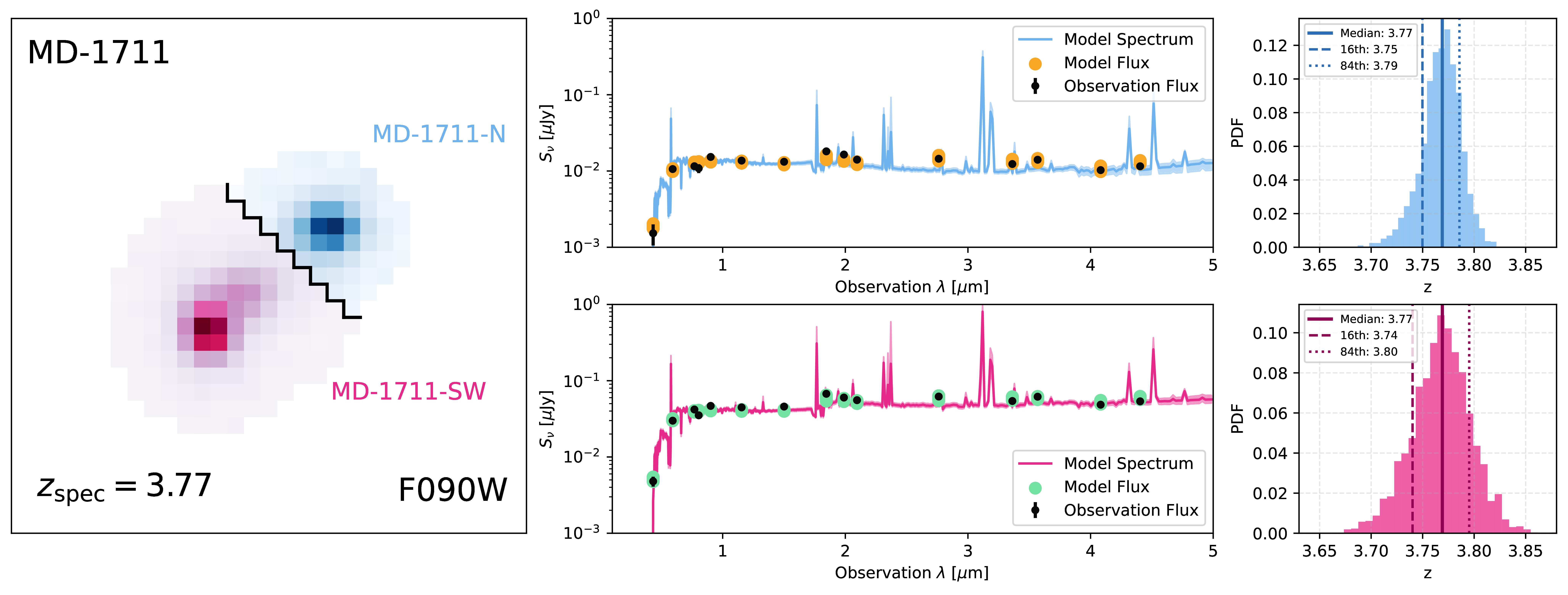} 
\caption{
Illustration of the SED fitting results for the LAE merger system MD-1711 at $z=3.77$. 
The symbols and layout follow the same convention as in Figure~\ref{fig:6666}.}
\label{fig:1711}
\end{figure*}

Figures~\ref{fig:6666}, \ref{fig:82}, and \ref{fig:1711} present the SED fitting results for each galaxy component in the LAE merger candidates, along with the corresponding redshift posterior distributions from \textsc{Bagpipes}. 
As shown in these figures and in Table~\ref{table2}, the photometric redshifts ($z_{\mathrm{phot}}$) of the individual components are mutually consistent and in excellent agreement with their spectroscopic counterparts. The normalized redshift offset, $\Delta z / (1 + z)$, satisfies $0 \leq \Delta z / (1 + z) \leq 0.02$, well within typical uncertainties for high-redshift photometric analyses. 

To further investigate the nature of these systems, we have evaluated several alternative interpretations. The hypothesis that the double-nucleus morphology arises from radiative transfer effects within a single galaxy is disfavored by the clear spatial separation ($0.15''$--$0.30''$) and the substantial stellar masses of the individual components—properties characteristic of distinct galaxies rather than star-forming clumps. Our source decomposition method (Section \ref{3.1}) was applied only to systems exhibiting unambiguous double-nucleus structures in multiple bands, thereby minimizing the risk of artificial splitting. The alternative scenario of a chance superposition of unrelated galaxies is ruled out by the tight agreement in photometric redshifts and the shared Ly$\alpha$ emission between the components. Collectively, the evidence from redshift consistency, stellar masses, and morphology tends to support the interpretation that these systems are genuine galaxy mergers.

\subsection{Resolved SED Analysis}\label{4.2}

We performed pixel-by-pixel SED fitting for each LAE merger system using the \textsc{Bagpipes} configuration described in Section~\ref{3.2}.  
For every pixel, we derived a set of physical parameters -- including stellar mass, mass-weighted age, dust attenuation ($A_V$), rest-frame UV slope ($\beta$), star formation rate (SFR), and specific star formation rate (sSFR) -- and mapped their spatial distributions. 
 
\begin{figure*}[t]
\centering
\includegraphics[width=1\textwidth]{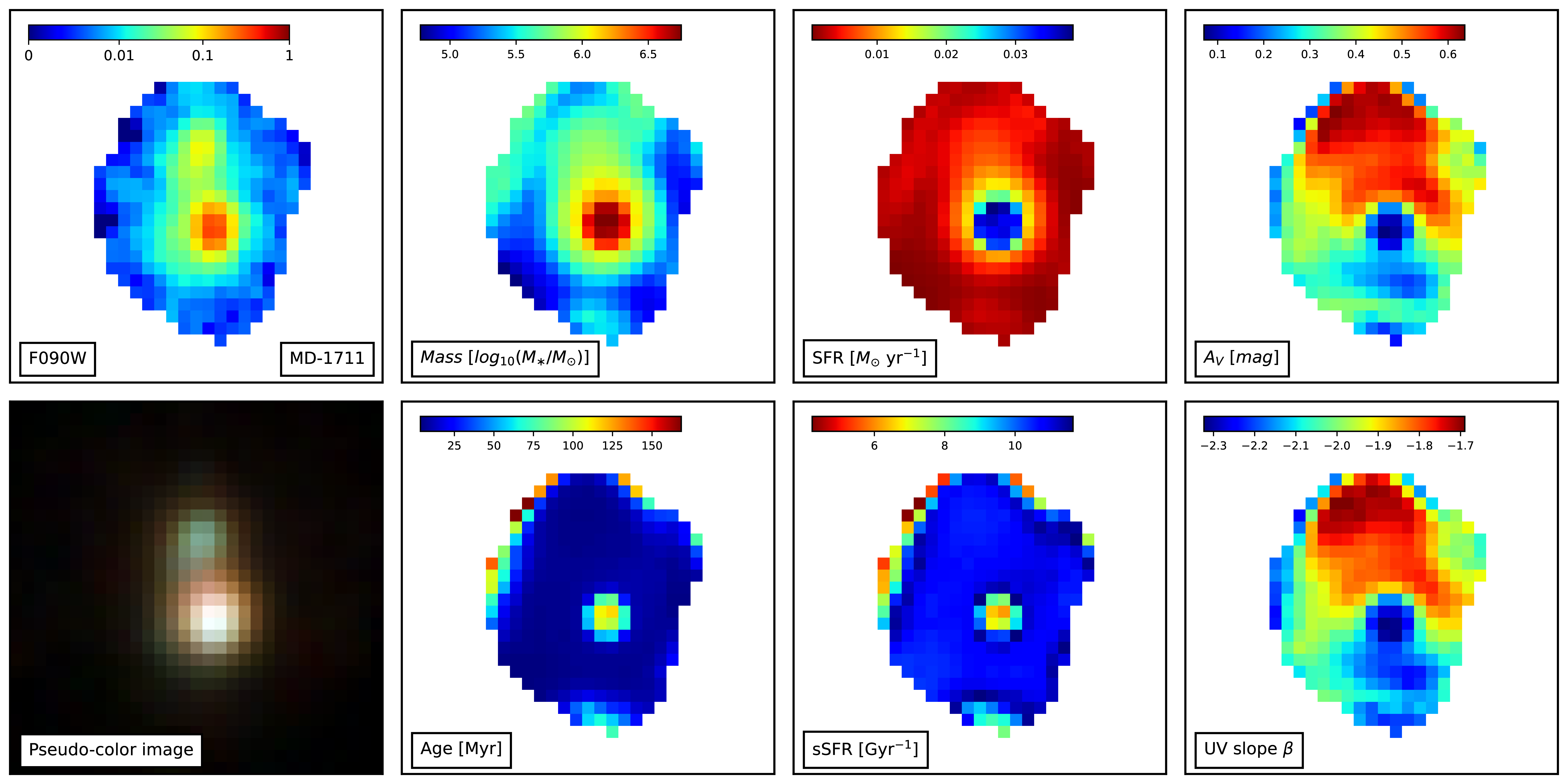} 
\caption{
Resolved pixel-by-pixel SED fitting results for the galaxy MD-6666 at $z = 3.43$. 
The top-left panel shows the JWST/NIRCam F090W image; the remaining panels display maps of stellar mass, mass-weighted age, dust attenuation ($A_V$), star formation rate (SFR), specific SFR (sSFR), pseudo-color image, and UV slope ($\beta$). 
Stellar mass, age, $A_V$, SFR, and sSFR are derived directly from \textsc{Bagpipes} fits, while $\beta$ is computed from the best-fit spectra produced by \textsc{Bagpipes}. 
Each pixel corresponds to $0.03''$, and all panels span $1'' \times 1''$. 
Note that the color scale is unique to each panel.
}
\label{fig:6666_sed}
\end{figure*}

\begin{figure*}[t]
\centering
\includegraphics[width=1\textwidth]{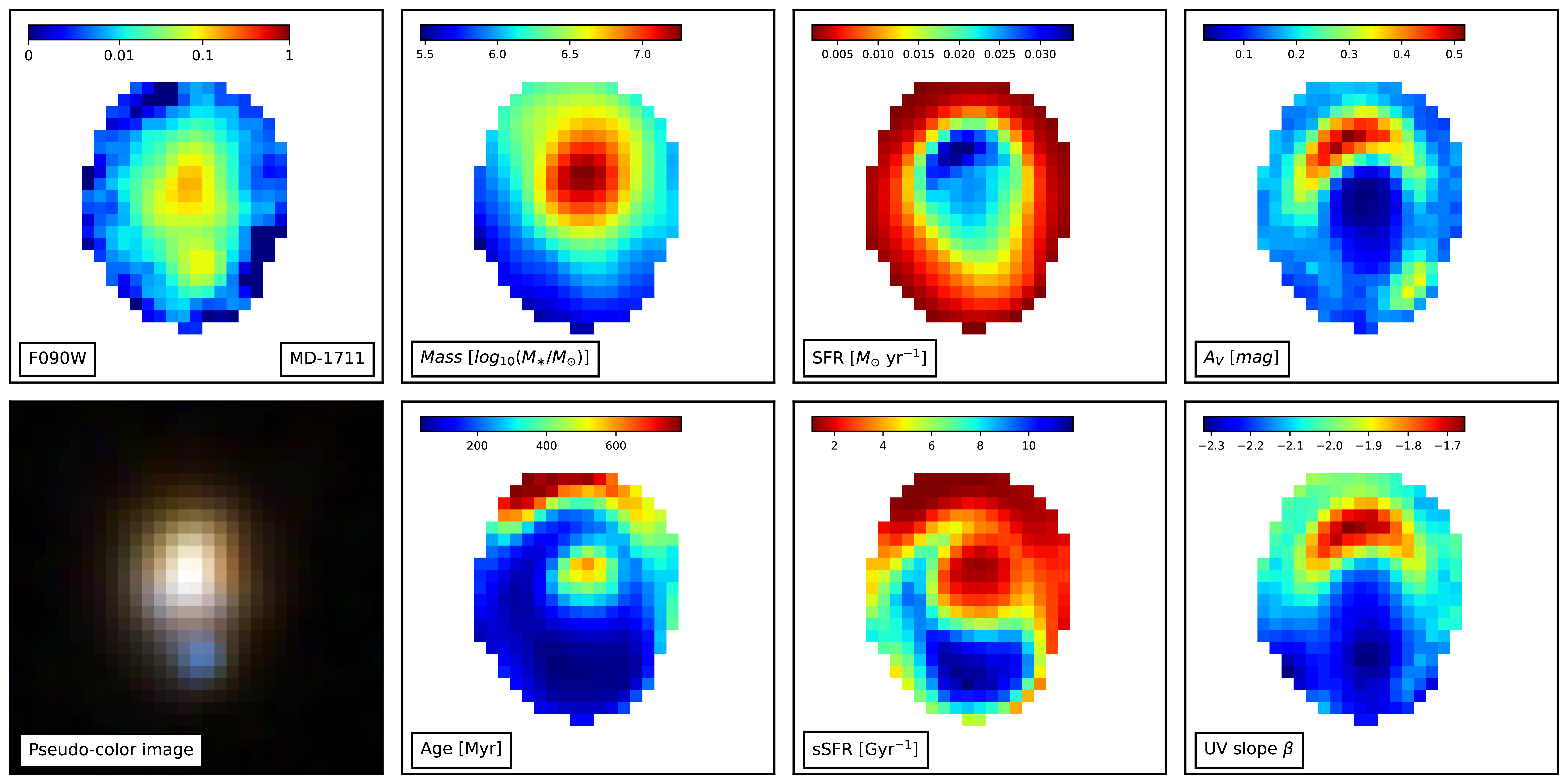} 
\caption{
Resolved pixel-by-pixel SED fitting results for the galaxy MD-82 at $z=3.61$. The symbols and layout follow the same convention as in Figure \ref{fig:6666_sed}.
}
\label{fig:82_sed}
\end{figure*}

\begin{figure*}[t]
\centering
\includegraphics[width=1\textwidth]{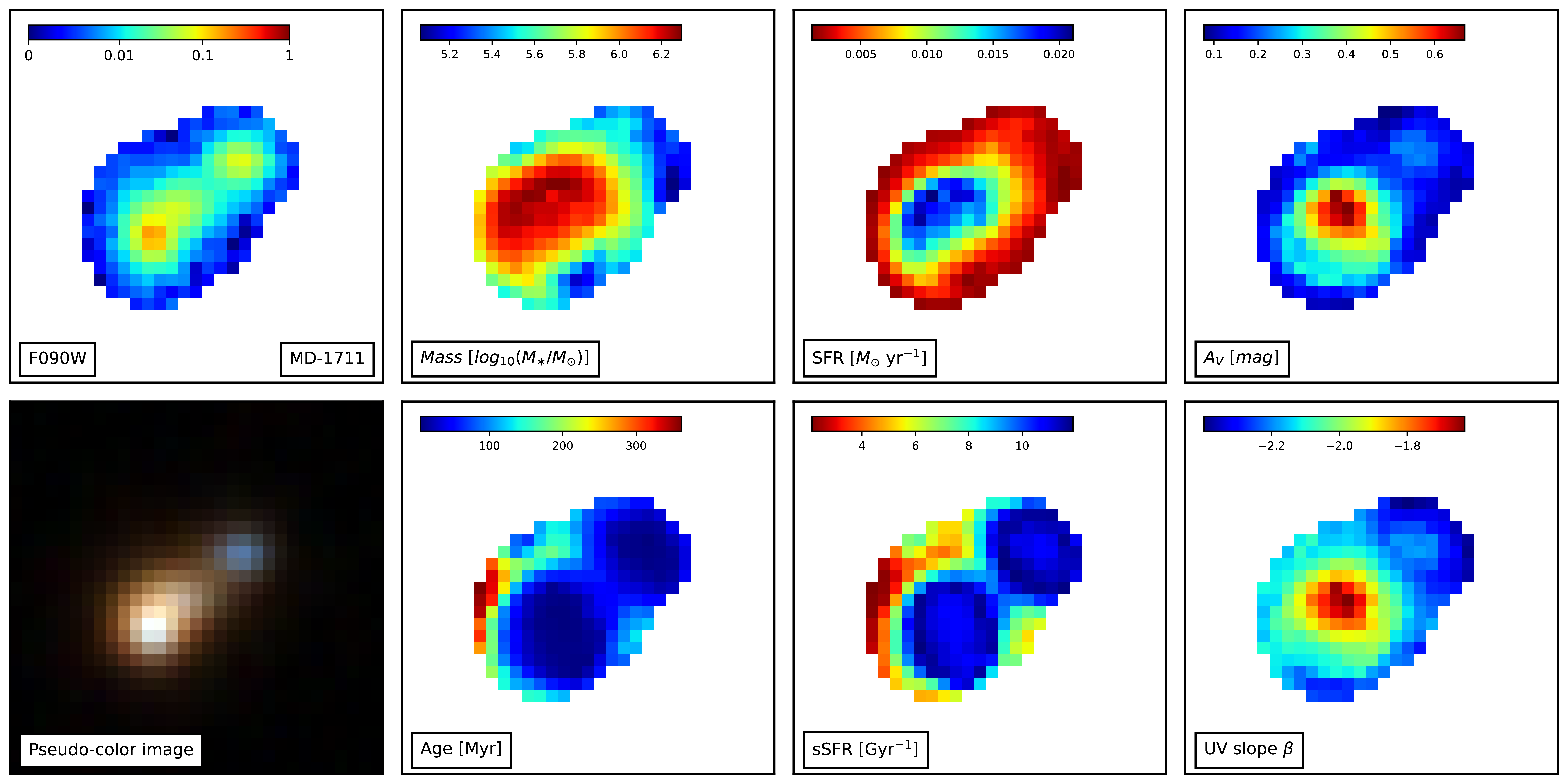} 
\caption{
Resolved pixel-by-pixel SED fitting results for the galaxy MD-1711 at $z=3.77$. The symbols and layout follow the same convention as in Figure \ref{fig:6666_sed}.
}
\label{fig:1711_sed}
\end{figure*}

Figures~\ref{fig:6666_sed}, \ref{fig:82_sed}, and \ref{fig:1711_sed} present the spatially resolved, pixel-by-pixel SED fitting results for the three LAE merger systems. 
In the following subsections, we analyze the results for each system individually.

\subsubsection{MD-6666}

Figure~\ref{fig:6666_sed} presents the results of our pixel-by-pixel SED fitting analysis for the LAE merger system MD-6666.  
The JWST/NIRCam F090W image reveals two distinct components: a brighter galaxy in the southern region and a fainter one to the north.  
As derived in Section~\ref{4.1}, the stellar mass ratio is MD-6666-N\,:\,MD-6666-S = 1\,:\,2, indicating that this system is undergoing a major merger. This interpretation is further confirmed by the stellar mass map from \textsc{Bagpipes}, which clearly shows that the southern component is more massive, while the northern galaxy is less massive and more spatially diffuse.

The SFR map shows that active star formation is concentrated in the central region of the southern (more massive) galaxy. 
In contrast, the sSFR exhibits a relatively uniform distribution across the system, with a modest decline toward the center of the southern component. This difference reflects the underlying disparity in stellar mass between the two merging galaxies.

Both the dust attenuation ($A_V$) and UV slope ($\beta$) maps reveal a striking spatial discontinuity aligned with the component axis. 
The $A_V$ distribution displays a pronounced north–south asymmetry: the northern galaxy exhibits higher dust attenuation ($A_V \gtrsim 0.5$~mag), whereas the southern component shows significantly lower attenuation ($A_V \lesssim 0.3$~mag). 
Correspondingly, the $\beta$ map shows a clear bimodal structure along the same direction, with redder UV slopes in the north ($\beta \simeq -1.7$) and markedly bluer slopes in the south ($\beta \simeq -2.3$). 
This behavior is naturally explained by dust attenuation: UV photons are preferentially absorbed at shorter wavelengths, resulting in redder observed spectra (i.e., higher $\beta$ values) in dustier regions. 
Thus, the spatial variation in $\beta$ is primarily governed by the distribution of dust.

Finally, regions with very young stellar ages ($\lesssim$ 30 Myr) closely coincide with areas of young stellar populations and elevated sSFR, indicative of vigorous ongoing star formation. 
This finding is consistent with recent observations by \citet{Lines25} and \citet{Clara23}.

\subsubsection{MD-82}

Figure~\ref{fig:82_sed} presents the pixel-by-pixel SED fitting results for the LAE merger system MD-82. 
The JWST/NIRCam F090W image reveals a close-pair configuration with a pronounced stellar mass asymmetry.  
As derived in Section~\ref{4.1}, the system has a stellar mass ratio of approximately 10\,:\,1, classifying it as a minor merger. 
This is confirmed by the \textsc{Bagpipes}-derived stellar mass map, which shows that the northern galaxy is significantly more massive than its southern counterpart. The color bar (spanning $\log(M_\ast/M_\odot) \approx 5.5$--$7.25$) further highlights this stark mass contrast.  
The SFR distribution closely traces the stellar mass, with the strongest activity concentrated in the central region of the northern galaxy.

Interestingly, the mass-weighted age and sSFR maps reveal two distinct sites of recent star formation: (i) the southern, less massive galaxy, which hosts younger stellar populations and elevated sSFR, and (ii) a prominent ring-like structure encircling the core of the massive northern galaxy. 

We interpret this morphology as characteristic of a late-stage merger, in which the primary galaxy is tidally disrupting its low-mass companion and accreting its gas. 
Gravitational compression of this infalling material along tidal streams--driven by the deep potential well of the primary--likely triggers the observed burst of star formation, giving rise to the ring-like enhancement in both sSFR.

Moreover, a localized enhancement in dust attenuation ($A_V$) and UV slope ($\beta$) is observed just north of the northern galaxy’s core. 
This feature may represent the residual signature of past star formation episodes, during which dust was enriched and retained in the interstellar medium. 
The spatial correlation between elevated $A_V$ and redder $\beta$ values further supports the interpretation that MD-82 is in a mature phase of its merger evolution.

\subsubsection{MD-1711}

Figure~\ref{fig:1711_sed} presents the results of our pixel-by-pixel SED fitting for MD-1711 -- the highest-redshift system in our sample -- performed with \textsc{Bagpipes}. 
As discussed in Section~\ref{4.1}, the stellar mass ratio between the two components is MD-1711-NE\,:\,MD-1711-SW = 1\,:\,10, consistent with a minor merger scenario. 
This interpretation is corroborated by both the JWST/NIRCam F090W imaging and the stellar mass map, which show the more massive galaxy located in the southwest and its lower-mass companion to the northeast.

Although no prominent tidal features are visible in the rest-frame optical imaging, the two galactic nuclei lie in close proximity, and a bridge-like structure connecting the components emerges clearly in the SFR and mass-weighted age maps. 
Both components exhibit moderate dust attenuation ($A_V \sim 0.3$--$0.6$~mag), suggesting that dust was already present prior to the interaction. Notably, the younger stellar ages in both galaxies are spatially coincident with elevated sSFR, directly implying the active star formation. This demonstrates that the ongoing merger has significantly influenced star formation activity across distinct structural components of the system.

We therefore interpret MD-1711 as being in a transitional merger phase. Despite the absence of extended tidal tails or a fully disrupted stellar body, the combination of (i) closely separated nuclei, (ii) a bridge traced by SFR and stellar age, and (iii) widespread young stellar populations.

\section{Discussion}\label{5}

\subsection{Integrated versus Spatially Resolved Stellar Mass}

\begin{figure*}[t]
\centering
\includegraphics[width=0.8\textwidth]{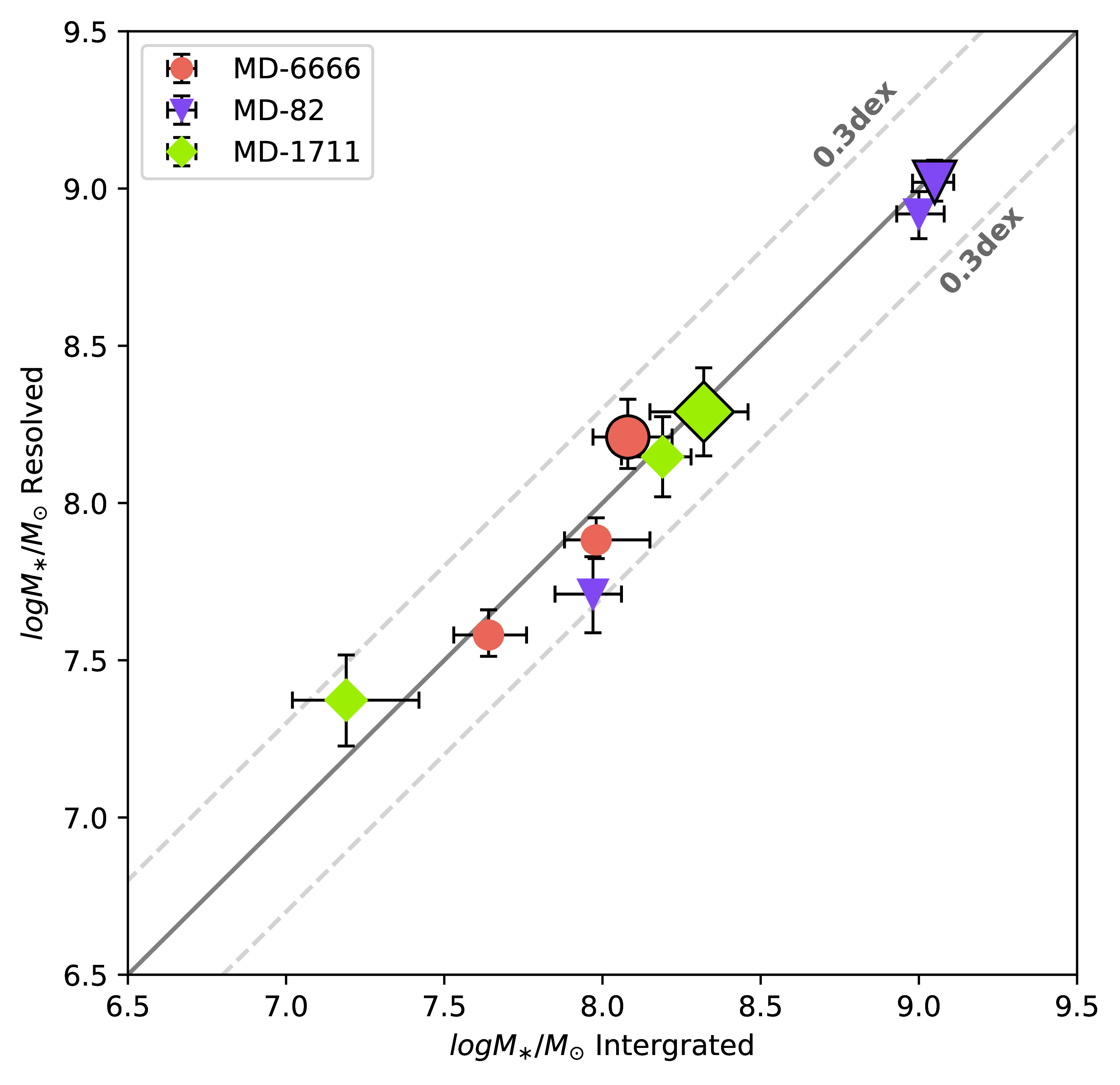} 
\caption{
Comparison between integrated stellar masses with the total spatially resolved stellar masses for both individual galaxy components and entire merger systems. 
Components belonging to the same system share the same color and marker: red circles for MD-6666 ($z_{\rm spec} = 3.43$), purple inverted triangles for MD-82 ($z_{\rm spec} = 3.61$), and green diamonds for MD-1711 ($z_{\rm spec} = 3.77$). 
Smaller symbols denote individual components; larger symbols represent the total stellar mass of the full system. 
Error bars show the 16th and 84th percentiles of the \textsc{Bagpipes} posterior distributions. 
The solid gray line indicates the one-to-one relation ($y = x$); dashed gray lines mark deviations of $\pm0.3$~dex.
}
\label{fig:mass_comparation}
\end{figure*}

In Section~\ref{4.1}, we performed individual SED fitting for each component of the LAE merger systems to obtain integrated (i.e., total) stellar mass estimates. 
In Section~\ref{4.2}, we conducted pixel-by-pixel SED fitting to construct spatially resolved stellar mass maps. 
To evaluate the consistency between these two approaches, this subsection presents a direct comparison of the integrated and spatially resolved stellar mass measurements. 
For the integrated analysis, we adopt the 50th percentile of the posterior distribution as the best-fit stellar mass, with the 16th and 84th percentiles defining the lower and upper $1\sigma$ uncertainties, respectively. 
For the resolved analysis, we first isolate pixels associated with each galaxy component. For each pixel, we similarly take the 50th percentile of its stellar mass posterior as the representative value, using the 16th and 84th percentiles as the corresponding uncertainty bounds. 
We then sum the pixel-level stellar masses within each component to derive the total resolved stellar mass, enabling a direct comparison with the integrated estimate.

Figure~\ref{fig:mass_comparation} compares the integrated stellar masses with the total spatially resolved stellar masses for both the entire system and individual galaxy components. 
The larger data points represent the total stellar mass of the entire merger system derived from both methods, while the smaller points correspond to the individual components. 
As shown, the two methods yield highly consistent results, with discrepancies typically below 0.3~dex. 
This agreement between the integrated and total spatially resolved stellar masses indicates that recent star formation does not significantly outshine the light from older stellar populations in these systems \citep{Clara23}.

\subsection{Merger-Driven Ly\texorpdfstring{$\alpha$}{α} Escape without Starburst Enhancement}

\begin{figure*}[t]
\centering
\includegraphics[width=1\textwidth]{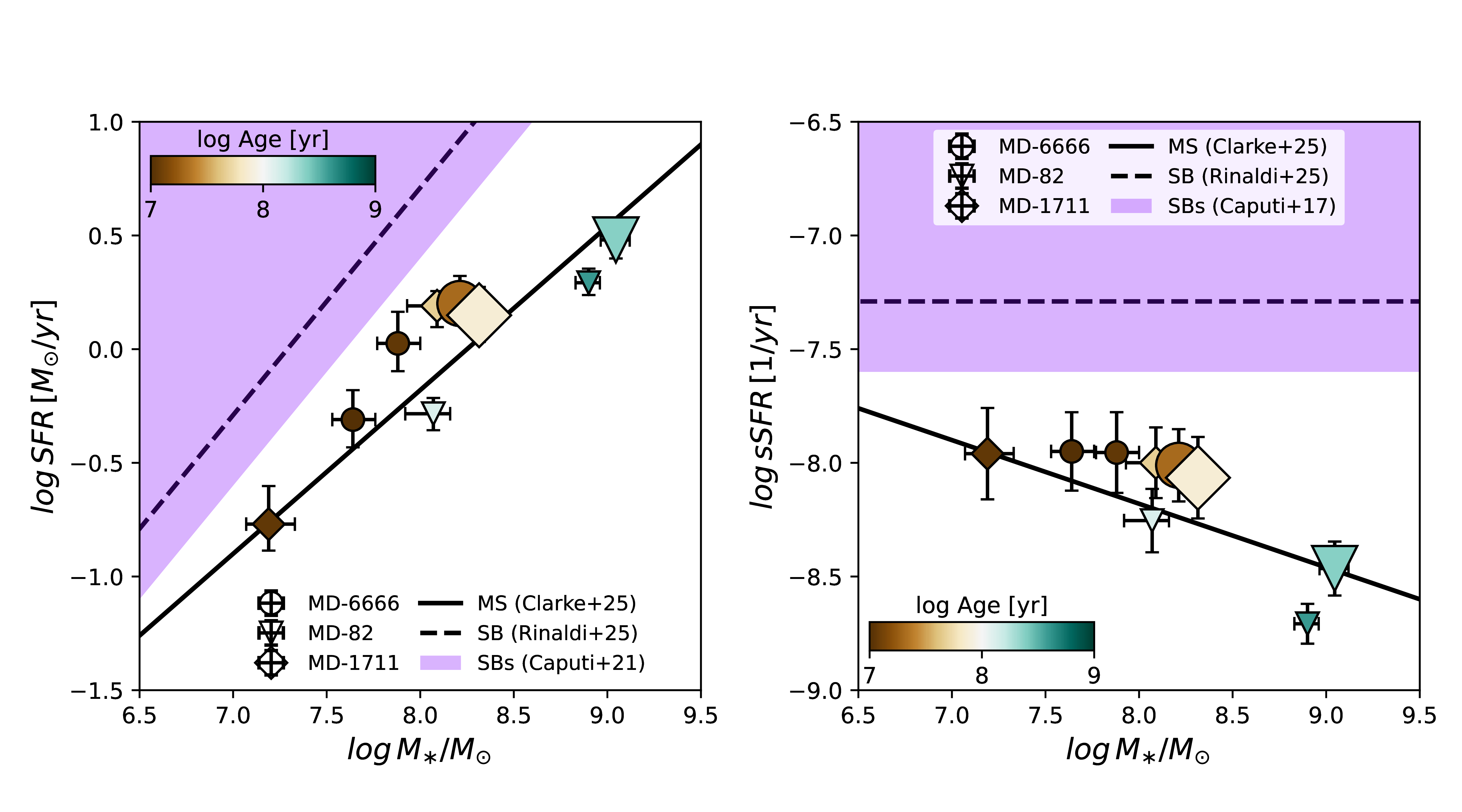} 
\caption{
Stellar mass versus SFR (left panel) and stellar mass versus specific SFR (sSFR; right panel). 
Symbol size distinguishes individual galaxy components (smaller symbols) from entire merger systems (larger symbols), while color encodes the mass-weighted stellar age. 
Different symbol shapes denote distinct LAE merger systems, consistent with Figure~\ref{fig:mass_comparation}. 
The purple shaded region marks the starburst regime ($\mathrm{sSFR} \geq 10^{-7.6}\,\mathrm{yr}^{-1}$; \citealt{Caputi17,Caputi21}). 
The solid black line shows the star-forming main sequence from \citet{Clarke2024ApJ...977..133C}, and the dashed black line indicates the starburst sequence from \citet{Rinaldi25}.
}
\label{fig:mass_sfr}
\end{figure*}

We investigate the relationship between star formation rate (SFR) and stellar mass using spatially resolved SED fitting results from \textsc{Bagpipes}. 
The total SFR for each system or component is obtained by summing the SFRs of individual pixels, following the same methodology used to estimate stellar masses. 
Uncertainties are quantified using the 16th and 84th percentiles of the posterior distributions.

Figure~\ref{fig:mass_sfr} presents our results for both the global properties of the LAE merger systems and their individual components. 
As shown, all data points lie on or near the star-forming main sequence, indicating that galaxy mergers do not significantly enhance star formation activity--at least in these systems. 
This is consistent with the findings of \citet{Ren25}, who also report that the majority of merging LAEs reside on the star-forming main sequence, in agreement with the ``Type II'' LAEs identified in the simulations of \citet{Shimizu10}. 
Recent observational and theoretical studies by \citet{Iani24} and \citet{Shimizu25} further propose a bimodal classification of LAEs based on stellar age. 
The properties of our LAE mergers closely resemble those of the older stellar population in their framework.

Our pixel-by-pixel maps of dust attenuation ($A_V$) and UV spectral slope ($\beta$) provide additional insight into the nature of Type II LAEs as envisioned in simulations: rather than triggering intense localized starbursts, mergers in these systems may facilitate Ly$\alpha$ photon escape primarily through the redistribution of gas and dust. 
This interpretation offers a plausible physical explanation for the spatial offset between Ly$\alpha$ emission peaks and UV continuum sources observed in Figure~\ref{fig:Sample_image}. 
It is worth noting, however, that this offset could alternatively arise from the resolution mismatch between MUSE integral-field spectroscopy and the higher-resolution JWST imaging \citep{Jiang24}.

\section{Summary}\label{6}

In this study, we selected three LAE merger systems at $3 < z < 4$ in the GOODS-S field, identified based on their multicomponent morphologies using deep MUSE integral-field spectroscopy and high-resolution \textit{JWST} imaging. The multi-wavelength photometric dataset comprises 15 bands -- 11 from \textit{JWST}/NIRCam and 4 from \textit{HST}/ACS -- spanning 0.4 to 5\,$\mu$m. 
We performed individual SED fitting for each component within the LAE merger systems, confirming that they are likely genuine physical mergers rather than chance projections, and that the smaller components represent distinct galaxies rather than star-forming clumps embedded in a single disk. 
Additionally, pixel-by-pixel SED fitting across each system reveals clear spatial gradients in key physical properties. 
The excellent agreement between integrated and spatially resolved stellar mass estimates -- for both individual components 
and the full systems -- supports a robust decomposition of the mergers and indicates widespread, 
distributed star formation throughout their structures. 
Analysis of the star formation rates shows that none of the systems exhibit strong starburst activity; 
instead, they lie firmly on the star-forming main sequence. 
By examining the MUSE-Deep narrowband data together with the spatial offsets between Ly$\alpha$ emission 
peaks and regions of low dust attenuation ($A_V$), we propose that merger-driven gas kinematics -- rather than enhanced 
star formation -- play the dominant role in facilitating Ly$\alpha$ photon escape. 
In future work, high-resolution near-infrared spectroscopy with \textit{JWST} will help further test and validate our results.

\begin{acknowledgements}
We gratefully acknowledge the valuable feedback from the reviewer and the editorial team, which has significantly contributed to improving the quality of this work.
This work is supported by the National Natural Science Foundation of China (NSFC  grants No. 12273052, 11733006, 12090040, 12090041, 12073051 and 12503015), the science research grants from  the China Manned Space Project (No.CMS-CSST-2021-A04). 
%
%
This work utilizes observational data acquired through the NASA/ESA/CSA James Webb Space Telescope, retrieved from the Mikulski Archive for Space Telescopes (MAST) maintained by the Association of Universities for Research in Astronomy, located at the Space Telescope Science Institute. 

\textit{Software}: Astropy \citep{astropy:2013,astropy:2018,astropy:2022}, 
Photoutils \citep{photutils24}, 
PyPHER \citep{pypher}, 
Bagpipes \citep{Bagpipes18}, 
Numpy \citep{numpy20}, 
Matplotlib \citep{matplotlib07}
\end{acknowledgements}

\bibliographystyle{raa}
\bibliography{ms}

\appendix                  

\clearpage  

\section{Multi-band imaging of three LAE merger systems} \label{sec:app1}

\begin{figure}[H]  
\centering
\includegraphics[width=1\textwidth]{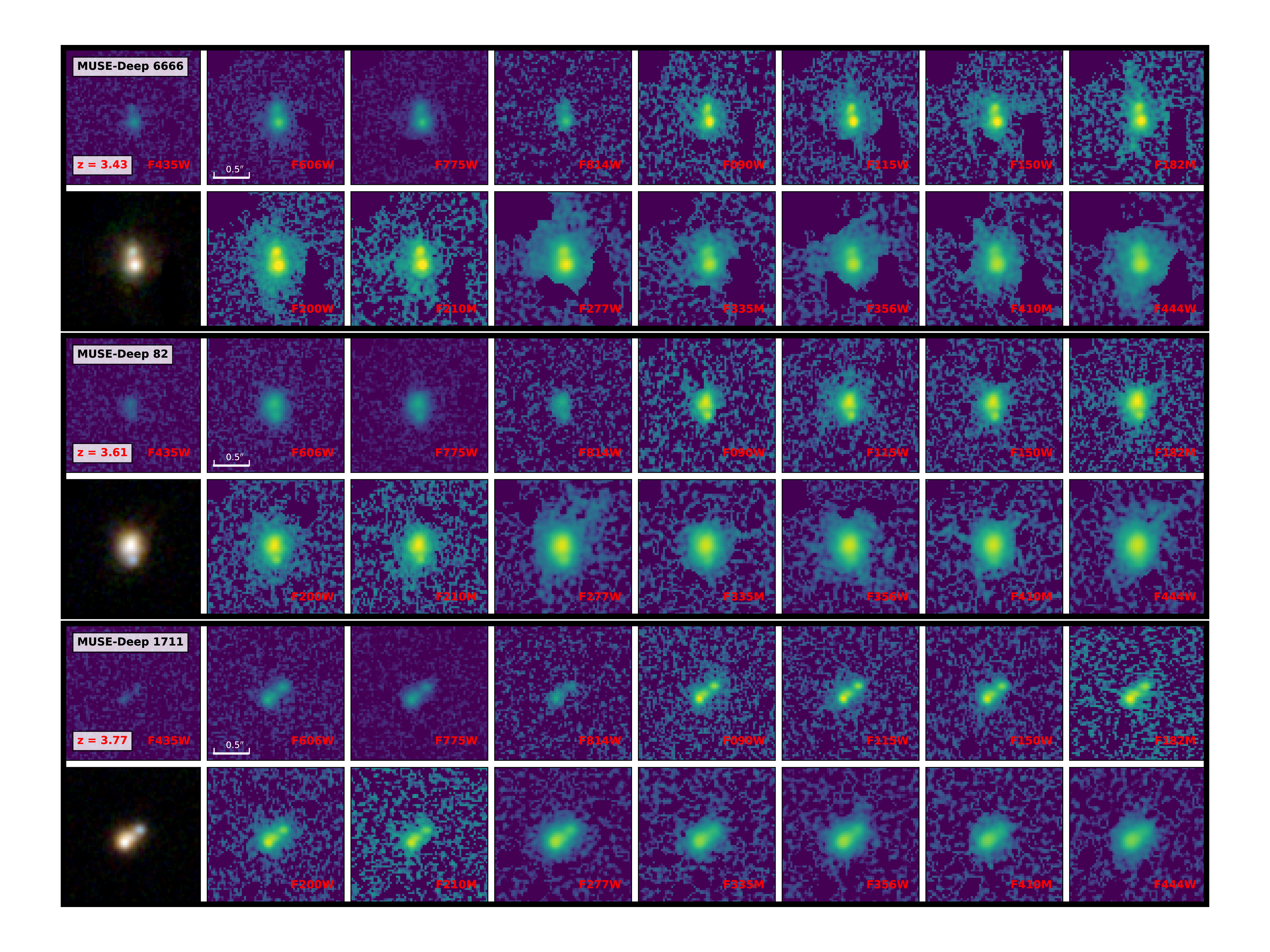} 
\caption{
Multi-band imaging data of the LAE merger systems are presented, with some images including masks. Each image spans $2'' \times 2''$. Pseudo-color composites were constructed using seven \textit{JWST}/NIRCam bands: blue (F090W, F115W, F150W), green (F200W, F277W), and red (F356W, F444W). The apparent differences from Figure~\ref{fig:Sample_image} are due to variations in dynamic range settings.
}
\label{fig:mul_image}
\end{figure}

We present high-resolution, multi-band imaging of our three LAE merger systems obtained with \textit{HST}/ACS and \textit{JWST}/NIRCam, including the pseudo-color composites shown in Figure~\ref{fig:mul_image}.  
The montage displays 15 bands in total: F435W, F606W, F775W, F814W, F090W, F115W, F150W, F182M, F200W, F210M, F277W, F335M, F356W, F410M, and F444W.

\clearpage  
\end{document}